\documentclass[preprint, 5p]{elsarticle}

\usepackage{amsmath,amssymb,bm,color}
\usepackage{lineno,hyperref}
\usepackage{multirow}
\usepackage{float}
\usepackage{booktabs}
\usepackage{array}
\usepackage{epstopdf}

\begin{document}

\begin{frontmatter}

\title{Missed prediction of the neutron halo in $^{37}$Mg}

\author[INPC,PKU]{K. Y. Zhang}

\author[ICA]{S. Q. Yang\corref{cor1}}
\ead{yangsiqi@caep.cn}

\author[BH]{J. L. An}

\author[BH]{S. S. Zhang\corref{cor1}}
\ead{zss76@buaa.edu.cn}

\author[RISP]{P. Papakonstantinou}
\author[SU]{M.-H. Mun}
\author[CENS]{Y. Kim}

\author[INPC,KLNP]{H. Yan}


\cortext[cor1]{Corresponding Author}

\address[INPC]{Institute of Nuclear Physics and Chemistry, CAEP, Mianyang, Sichuan 621900, China}
\address[PKU]{State Key Laboratory of Nuclear Physics and Technology, School of Physics, Peking University, Beijing 100871, China}
\address[ICA]{Institute of Computer Application, CAEP, Mianyang, Sichuan 621900, China}
\address[BH]{School of Physics, Beihang University, Beijing 100191, China}
\address[RISP]{Rare Isotope Science Project, Institute for Basic Science, Daejeon 34000, Korea}
\address[SU]{Department of Physics and Origin of Matter and Evolution of Galaxy Institute, Soongsil University, Seoul 06978, Korea}
\address[CENS]{Center for Exotic Nuclear Studies, Institute for Basic Science, Daejeon 34126, Korea}
\address[KLNP]{Key Laboratory of Neutron Physics, Institute of Nuclear Physics and Chemistry, CAEP, Mianyang, Sichuan 621900, China}

\begin{abstract}
  Halo phenomena have long been an important frontier in both experimental and theoretical nuclear physics. $^{37}$Mg was identified as a halo nucleus in 2014 and remains the heaviest nuclear halo system to date. While the halo phenomenon in $^{37}$Mg was not predicted before the discovery, its description has been still challenging afterwards. In this Letter, we report a microscopic and self-consistent description of the neutron halo in $^{37}$Mg using the deformed relativistic Hartree-Bogoliubov theory in continuum (DRHBc) that was developed in 2010. The experimental neutron separation energies and empirical matter radii of neutron-rich magnesium isotopes as well as the deformed $p$-wave halo characteristics of $^{37}$Mg are well reproduced without any free parameters. In particular, the orbital occupied by the halo neutron in $^{37}$Mg, exhibiting $p$-wave components comparable to those suggested in experiments, remains consistent across various employed density functionals including PC-F1, PC-PK1, NL3*, and PK1. The DRHBc theory investigated only even-even magnesium isotopes in previous works and for that reason missed predicting $^{37}$Mg as a halo nucleus before 2014. Although the core and the halo of $^{37}$Mg are both prolate, higher-order shape decoupling on the hexadecapole and hexacontatetrapole levels is predicted.
\end{abstract}

\end{frontmatter}


\section{Introduction}

Since the discovery of the first halo nucleus, $^{11}$Li \cite{Tanihata1985PRL}, the investigation of halo phenomena in exotic nuclei far from the $\beta$ stability line has been a major driving force in the development of radioactive ion beam facilities worldwide.
Over nearly forty years of exploration, 13 halo nuclei have been identified, and 10 candidates have been suggested in experiments (see Fig.~1.4 of Ref. \cite{Tanihata2013PPNP} and Fig.~1 of Ref. \cite{Zhang2023PRC(L)}).
While all these identified and candidate halo nuclei locate in the light-mass region of the nuclear chart, many theoretical efforts have been made to explore halos in medium-mass and heavy nuclei \cite{Rotival2009PRC1,Rotival2009PRC2,Meng2015JPG}.
The theoretical approaches devoted to the description and prediction of halo phenomena include the few-body model \cite{Zhukov1993PhysRep,Hansen1995ARNPS}, shell model \cite{Otsuka1993PRL, Kuo1997PRL}, antisymmetrized molecular dynamics (AMD) \cite{Horiuchi1994ZPA, Itagaki1999PRC}, halo effective field theory \cite{Ryberg2014PRC, Ji2014PRC}, nonrelativistic \cite{Terasaki1996NPA, Schunck2008PRC} and relativistic \cite{Meng1996PRL, Meng1998PRL} density functional theories, \emph{ab initio} calculations within the no-core shell model with continuum \cite{Calci2016PRL}, etc.

$^{37}$Mg, identified as a deformed $p$-wave halo nucleus by Kobayashi \emph{et al.} \cite{Kobayashi2014PRL} and Takechi \emph{et al.} \cite{Takechi2014PRC(R)} in 2014, is the heaviest nuclear halo system to date.
This neutron-rich nucleus was first discovered by Sakurai \emph{et al.} \cite{Sakurai1996PRC(R)} in 1996.
Between 1996 and 2014, $^{37}$Mg was involved in several theoretical studies \cite{Ren1996PLB,Chen2005ChinPhy,Zhi2006PLB,Hamamoto2007PRC,Horiuchi2012PRC,Sharma2013IJMPE}.
However, its possible halo structure was not indicated during this period.
For instance, in the relativistic mean field (RMF) calculations, a monotonic and smooth increase of the neutron skin thickness with the neutron number in magnesium isotopes was found \cite{Ren1996PLB}.
Although it was suggested that $^{37}$Mg may prefer being deformed due to the Jahn-Teller effect, important $p$-wave resonant levels were not obtained based on Woods-Saxon potentials \cite{Hamamoto2007PRC}.
The enhancement of the reaction cross section for $^{37}$Mg was not predicted by the Glauber model, using the nuclear densities from either the Skyrme Hartree-Fock or the RMF theory \cite{Horiuchi2012PRC,Sharma2013IJMPE}.

Even after the discovery of the halo in $^{37}$Mg, achieving a microscopic and self-consistent description of this nucleus has remained challenging.
Based on quadrupole-deformed Woods-Saxon potentials, the analytical continuation of the coupling constant method \cite{Xu2015PRC}, the complex momentum representation method \cite{Fang2017PRC}, and the Green's function method \cite{Sun2020PRC} were applied to solve the Dirac equation for $^{37}$Mg.
It turns out that the $p$-wave configuration becomes energetically favored only when the deformation parameter $\beta_2\gtrsim0.5$.
Similarly, by invoking a quadrupole deformation parameter of 0.5 and further fitting the depth parameter to the neutron separation energy, the Woods-Saxon potential based Hartree-Fock-Bogoliubov (HFB) calculations could provide reasonable nuclear densities for the Glauber model to reproduce the reaction cross section for $^{37}$Mg \cite{Urata2017PRC}.
Another similar attempt in the Glauber model was to significantly increase the diffuseness parameter of the Gaussian-form density distribution, which was fitted from the RMF calculations for $^{37}$Mg \cite{Sharma2016PRC}.
Microscopic AMD calculations successfully reproduced the deduced matter radii of $^{24\text{--}36}$Mg but considerably underestimated that of $^{37}$Mg \cite{Watanabe2014PRC,Choudhary2021PRC}.
The results from microscopic HFB calculations were density-functional dependent, leading to diverse predictions for the ground state of $^{37}$Mg.
For example, SLy4 and UNEDF1 density functionals predicted that the valence neutron occupies a $5/2^-$ orbital without any $p$-wave components \cite{Xiong2016CPC,Kasuya2020PTEP}, whereas the M3Y-P6 functional suggested that it occupies a $1/2^-$ orbital with $p$-wave components \cite{Nakada2018PRC(R)}.

In this Letter, we report a microscopic and self-consistent description of the deformed $p$-wave halo characteristics of $^{37}$Mg using the deformed relativistic Hartree-Bogoliubov theory in continuum (DRHBc) without any free parameters.
In particular, it remains consistent across various employed density functionals that the halo neutron in $^{37}$Mg occupies a $1/2^-$ orbital with $p$-wave components comparable to those suggested in experiments.

\section{The DRHBc theory}

In 2010, Zhou \emph{et al}. \cite{Zhou2010PRC(R)} developed the DRHBc theory, which incorporates self-consistently the axial deformation, pairing correlations, and continuum effects, making it suitable for describing deformed halo nuclei.
By applying this theory to even-even magnesium isotopes, they predicted $^{42,44}$Mg as deformed halo nuclei with novel shape decoupling \cite{Zhou2010PRC(R),Li2012PRC}.
Later in 2012, the DRHBc theory was further extended to include the blocking effects in odd-mass or odd-odd nuclei \cite{Li2012CPL}, as well as a version with density-dependent meson-nucleon couplings \cite{Chen2012PRC}.
Although magnesium isotopes heavier than $^{40}$Mg are yet to be produced in the laboratory \cite{Baumann2007Nature}, the DRHBc theory has demonstrated its capabilities through various applications, e.g., successfully describing halo nuclei $^{17,19}$B \cite{Yang2021PRL,Sun2021PRC(1)}, $^{15,19,22}$C \cite{Sun2018PLB,Sun2020NPA}, and $^{31}$Ne \cite{Zhong2022SciChina}, predicting the $N = 28$ shell collapse and deformed halo in the recently discovered new isotope, $^{39}$Na \cite{Zhang2023PRC(L)}, revealing the connection between halo phenomena and nucleons in the classically forbidden region \cite{Zhang2019PRC}, investigating the shape coexistence from light to heavy nuclei \cite{In2020JKPS,Choi2022PRC,Kim2022PRC} and the prolate-shape dominance \cite{Guo2023PRC}, exploring the limits of nuclear existence \cite{Zhang2021PRC(L),Pan2021PRC,He2021CPC,In2021IJMPE}, and studying the rotational excitations of exotic nuclei and beyond mean-field correlations using the extension with angular momentum projection \cite{Sun2021SciBull,Sun2021PRC(2)} and collective Hamiltonian \cite{Sun2022CPC}.
Recently, efforts have been made to construct a DRHBc mass table incorporating both deformation and continuum effects \cite{Zhang2020PRC,Pan2022PRC,Zhang2022ADNDT}.

The DRHBc theory with meson-exchange and point-coupling density functionals has been introduced in details in Refs. \cite{Li2012PRC} and \cite{Zhang2020PRC}, respectively.
Here, we briefly present its formalism.
The relativistic Hartree-Bogoliubov (RHB) equations for nucleons \cite{Kucharek1991ZPA}
\begin{equation}\label{RHB}
\left(\begin{matrix}
h_D-\lambda & \Delta \\
-\Delta^* &-h_D^*+\lambda
\end{matrix}\right)\left(\begin{matrix}
U_k\\
V_k
\end{matrix}\right)=E_k\left(\begin{matrix}
U_k\\
V_k
\end{matrix}\right),
\end{equation}
are solved in a Dirac Woods-Saxon basis \cite{Zhou2003PRC,Zhang2022PRC}, whose wave function has an asymptotic behavior suitable for the description of weakly bound nuclei.
In Eq.~(\ref{RHB}), $h_D$ is the Dirac Hamiltonian, $\lambda$ the Fermi energy, $\Delta$ the pairing potential, and $E_k$ and $(U_k, V_k)^{\rm T}$ respectively the quasiparticle energy and wave function.
In the Dirac Hamiltonian,
\begin{equation}
      h_D(\bm{r})=\bm{\alpha}\cdot\bm{p}+V(\bm{r})+\beta[M+S(\bm{r})],
\end{equation}
the scalar potential $S(\bm r)$  and vector potential $V(\bm r)$ are expanded in terms of the Legendre polynomials,
\begin{equation}\label{legendre}
f(\bm r)=\sum_\lambda f_\lambda(r)P_\lambda(\cos\theta),~~\lambda=0,2,4,\cdots,
\end{equation}
so are various densities and the pairing potential,
\begin{equation}\label{Delta}
\Delta(\bm r_1,\bm r_2) = V^{\mathrm{pp}}(\bm r_1,\bm r_2)\kappa(\bm r_1,\bm r_2),
\end{equation}
where $V^{\mathrm{pp}}$ is the paring force and $\kappa$ the paring tensor \cite{Peter1980Book}.
A density-dependent pairing force of zero range,
\begin{equation}\label{pair}
      V^{\mathrm{pp}}(\bm r_1,\bm r_2)
  = V_0 \frac{1}{2}(1-P^\sigma)\delta(\bm r_1-\bm r_2)\left(1-\frac{\rho(\bm r_1)}{\rho_{\mathrm{sat}}}\right),
\end{equation}
which is widely used in the study of nuclear halos \cite{Meng2015JPG,Meng1998NPA,Meng1998PRC,Meng2006PPNP,Xiang2023Symmetry}, is adopted in the present DRHBc theory.
The work to implement a finite-range pairing force, e.g. the Gogny \cite{Meng1998NPA} or separable one \cite{Tian2009PLB}, in the DRHBc theory is in progress.

\section{Numerical Details}

The pairing strength $V_0=-342.5~\mathrm{MeV~fm}^3$ and the saturation density $\rho_{\mathrm{sat}}=0.152~\mathrm{fm}^{-3}$ in Eq.~(\ref{pair}), and a pairing window of $100$ MeV is chosen.
These can reproduce well the odd-even mass differences of Ca, Sn, Pb, and U isotopes as well as $N=20$ and $50$ isotones \cite{Xia2018ADNDT}.
An energy cutoff $E^+_{\mathrm{cut}}=300$ MeV and an angular momentum cutoff $J_{\max}=19/2~\hbar$ are adopted for the Dirac Woods-Saxon basis, which have been verified to yield converged results \cite{Zhang2020PRC}.
In Eq.~(\ref{legendre}), the Legendre expansion is truncated at $\lambda_{\max}=6$ \cite{Pan2019IJMPE,Zhang2020PRC}.
The blocking effects in odd-mass magnesium isotopes are taken into account via the equal filling approximation \cite{Perez-Martin2008PRC,Li2012CPL,Pan2022PRC}.

In our DRHBc calculations, we utilized four density functionals: PC-F1 \cite{Burvenich2002PRC}, PC-PK1 \cite{Zhao2010PRC}, NL3* \cite{Lalazissis2009PLB}, and PK1 \cite{Long2004PRC}.
Among these functionals, PC-F1 provides the best description of neutron separation energies.
For $^{35}$Mg, the PC-F1 calculated one-neutron separation energy, $S_n = 0.61$ MeV, reproduces the experimental value of $0.75^{+0.27}_{-0.27}$ MeV \cite{AME2020(3)}.
Regarding $^{37}$Mg, the PC-F1 calculated $S_n$ of $0.32$ MeV is in good agreement with both the estimation by Kobayashi \emph{et al}., $0.22^{+0.12}_{-0.09}$ MeV \cite{Kobayashi2014PRL}, and the latest atomic mass evaluation AME2020, $0.24^{+0.11}_{-0.11}$ MeV \cite{AME2020(3)}.
As for $^{39}$Mg that is likely unbound according to its empirical $S_n = -0.63^{+0.10}_{-0.10}$ MeV \cite{AME2020(3)}, PC-F1 also predicts a negative $S_n$ of $-0.92$ MeV, indicating $^{37}$Mg as the last bound odd-mass magnesium isotope.
Compared with PC-F1, although other density functionals yield slightly larger values for $S_n$, they predict similar bulk properties, density distributions, and single-particle levels for $^{37}$Mg.
In particular, all the employed functionals consistently indicate that the valance neutron of $^{37}$Mg occupies the $1/2^-$ orbital in the ground state, with quantitatively similar components for this orbital.
Therefore, the following results will be presented in detail based on the PC-F1 functional.

\section{Results and discussions}\label{Sec4}

\begin{figure*}[htbp]
  \centering
  \includegraphics[width=1.0\textwidth]{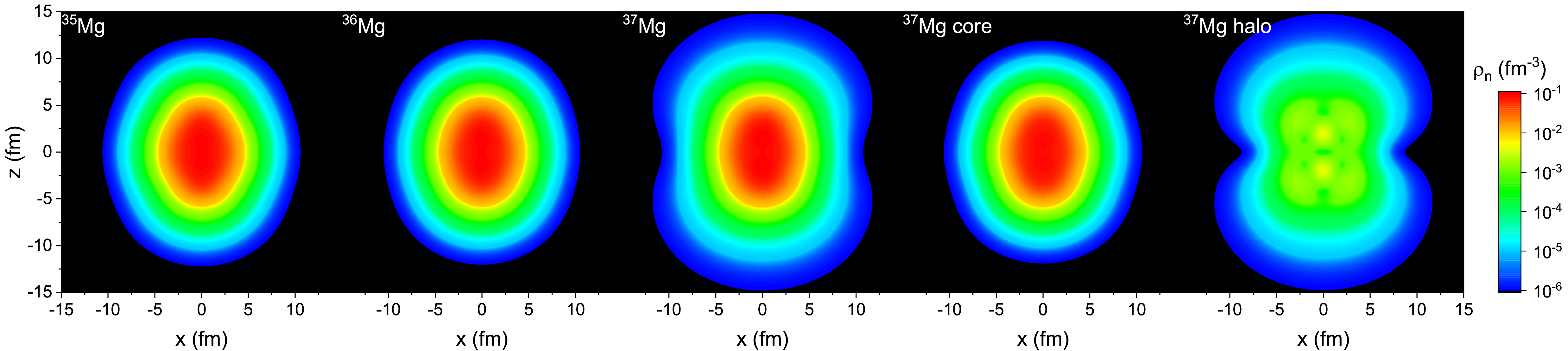}
  \caption{Two-dimensional neutron density distributions for $^{35\text{--}37}$Mg as well as $^{37}$Mg's core and halo in the $xz$ plane with the $z$ axis as the symmetry axis.}
\label{fig1}
\end{figure*}

The two-dimensional neutron density distributions of $^{35\text{--}37}$Mg are displayed in Fig.~\ref{fig1}.
Notably, the density distribution of $^{37}$Mg exhibits a significantly larger spatial extension compared to $^{35,36}$Mg, supporting its halo structure.
Figure~\ref{fig1} also presents the densities decomposed into the contribution of the core and the halo for $^{37}$Mg, which will be discussed below.
The root-mean-square (rms) matter radii calculated from nuclear densities for $^{35\text{--}37}$Mg are respectively 3.44, 3.49, and 3.57 fm, in satisfactory agreement with the empirical values of 3.44(03), 3.49(01), and 3.62(03) \cite{Watanabe2014PRC}.
It is worth noting that these empirical values are deduced from the measured reaction cross section by fitting it with the Woods-Saxon potential based calculations \cite{Watanabe2014PRC}.
Consequently, they depend on the choice of the depth and deformation parameters.
The ground-state deformation of $^{37}$Mg from microscopic DRHBc calculations is $\beta_2= 0.46$, slightly smaller than the deformation parameters of $\beta_2 \gtrsim 0.5$ suggested by the calculations based on Woods-Saxon potentials \cite{Xu2015PRC,Sun2020PRC,Urata2017PRC,Watanabe2014PRC}.
The DRHBc calculated densities can also be fed as microscopic inputs into the Glauber model to study the observables in nuclear reaction \cite{Zhong2022SciChina}.
Such a work on magnesium isotopes is in progress, and the preliminary results reasonably reproduce the existing experimental data for total cross sections and the parallel momentum distributions of reaction residues \cite{An2023}.

\begin{figure}[htbp]
  \centering
  \includegraphics[width=0.35\textwidth]{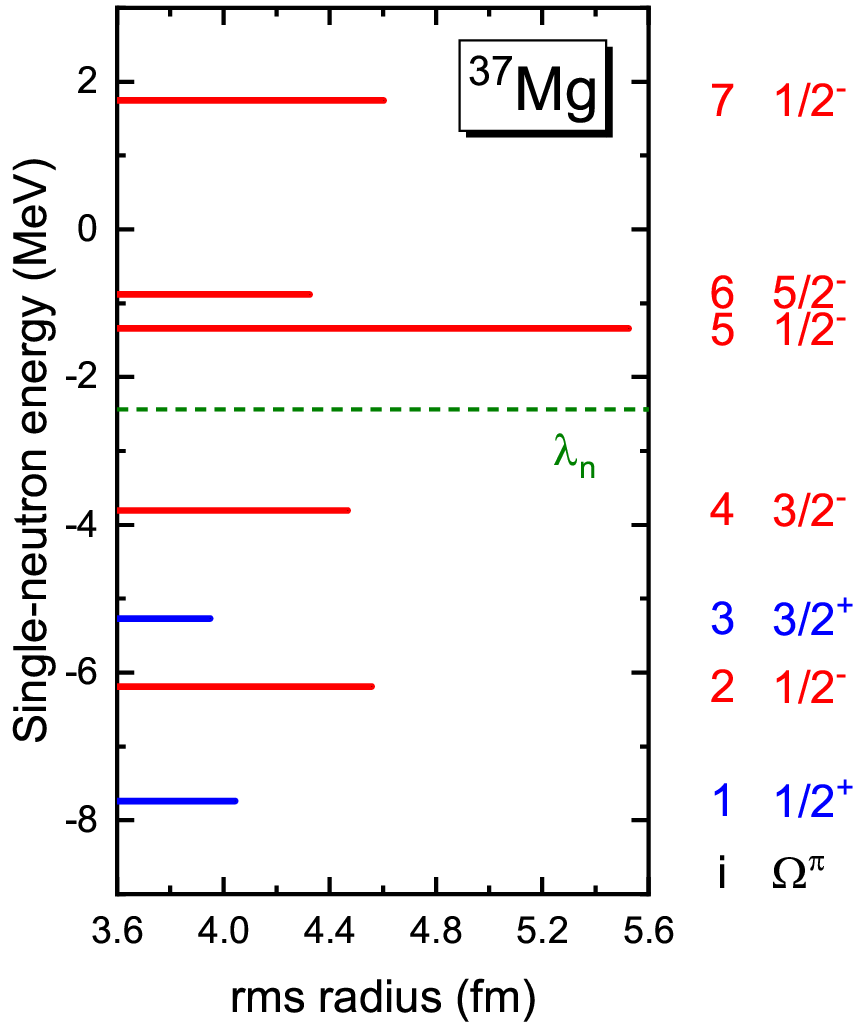}
  \caption{Energy versus rms radius for single-neutron orbitals around the Fermi energy $\lambda_n$ (dashed line) in the canonical basis for $^{37}$Mg. The order $i$ and quantum numbers $\Omega^\pi$ are given on the right. The occupation probabilities and main components of these orbitals are given in Table \ref{compon}.}
\label{fig2}
\end{figure}

\begin{table*}[htbp]
    \centering
    \caption{Occupation probabilities $v^2$ and main components of the single-neutron orbitals shown in Fig.~\ref{fig2}.}
    \begin{tabular}{c|c|c|ccccccc}
    \hline
    \hline
        \multirow{2}*{i} & \multirow{2}*{$\Omega^\pi$} & \multirow{2}*{$v^2$} & \multicolumn{7}{c}{Components} \\ \cline{4-10}
         &  & & $2s_{1/2}$ & $1d_{3/2}$ & $1d_{5/2}$ & $2p_{1/2}$ & $2p_{3/2}$ & $1f_{5/2}$ & $1f_{7/2}$  \\ \hline
        1 & $1/2^+$ & 0.99 & $57\%$ & $33\%$ & $4\%$ & - & - & - & -\\
        2 & $1/2^-$ & 0.98 & - & - & - & $4\%$ & $32\%$ & $2\%$ & $53\%$\\
        3 & $3/2^+$ & 0.96 & - & $90\%$ & $8\%$ & - & - & - & -\\
        4 & $3/2^-$ & 0.88 & - & - & - & - & $14\%$ & $4\%$ & $77\%$ \\
        5 & $1/2^-$ & 0.50 & - & - & - & {\color{red}\bm{$38\%$}} & {\color{red}\bm{$19\%$}} & $6\%$ & $32\%$\\
        6 & $5/2^-$ & 0.11 & - & - & - & - & - & $4\%$ & $93\%$ \\
        7 & $1/2^-$ & 0.02 & - & - & - & $2\%$ & $21\%$ & $63\%$ & $3\%$ \\ \hline
        \hline
    \end{tabular}
\label{compon}
\end{table*}

It is well recognized that weakly bound or continuum orbitals with low-$\ell$ components play a crucial role in the formation of nuclear halos \cite{Tanihata2013PPNP,Meng2015JPG,Hansen1995ARNPS,Meng2006PPNP}.
To understand the neutron halo in $^{37}$Mg, we present the single-neutron orbitals around the Fermi energy and their rms radii in Fig.~\ref{fig2}.
These orbitals are labeled by order $i$ and quantum numbers $\Omega^\pi$, where $\pi$ is the parity and $\Omega$ the projection of angular momentum on the symmetry axis.
Their occupation probabilities and main components are tabulated in Table \ref{compon}.
In Fig.~\ref{fig2}, the rms radius $\approx5.6$~fm of orbital 5 ($\Omega^\pi = 1/2^-$) is remarkably larger than the others.
It even far exceeds the rms matter radius, 3.57~fm, of the whole nucleus.
This can be understood from the components of its wave function.
As seen in Table~\ref{compon}, orbital 5 is occupied by the valence neutron of $^{37}$Mg, with $38\%$ $2p_{1/2}$ and $19\%$ $2p_{3/2}$ components, which are comparable to the $\approx 40\%$ $p$-wave neutron halo components suggested in experiments \cite{Kobayashi2014PRL}.
The low centrifugal barrier for $p$ wave allows considerable tunneling of the wave function into the classically forbidden region, which, together with the weak binding of orbital 5, accounts for its largest rms radius.
While certain low-$\ell$ components are found for orbitals 1 and 2, their rms radii are suppressed by the deep binding.
On the other hand, orbitals 6 and 7, despite being weakly bound or even in the continuum, play marginal roles due to their small occupation probabilities.

\begin{figure}[htbp]
  \centering
  \includegraphics[width=0.4\textwidth]{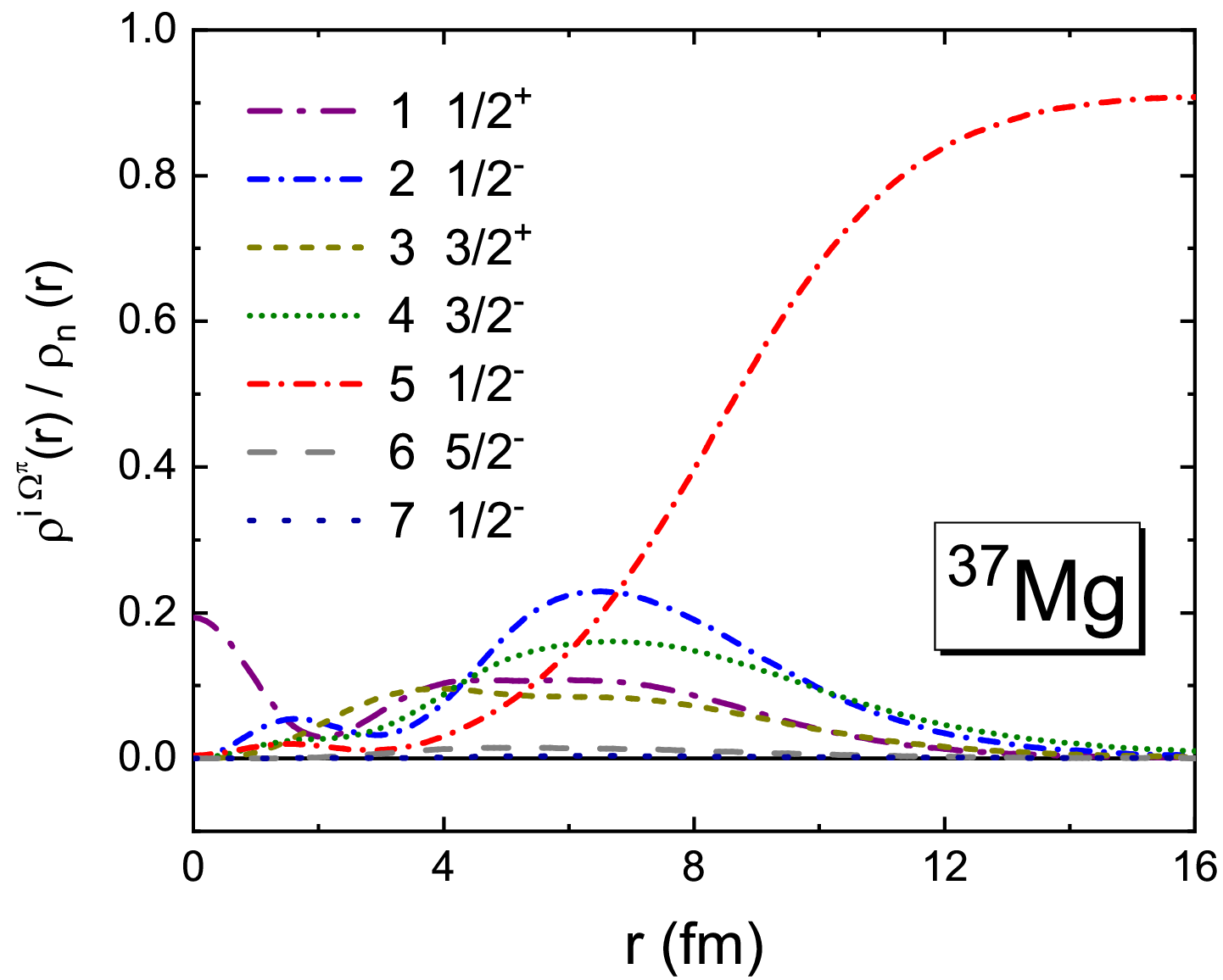}
  \caption{The contribution of each single-neutron orbital around the Fermi energy to the total neutron density as a function of the radial coordinate $r$.}
\label{fig3}
\end{figure}

To quantitatively analyze the role of orbital 5 in the formation of the halo, we show in Fig.~\ref{fig3} the contribution of each orbital around the Fermi energy to the total neutron density.
As observed, the contribution of orbital 5 rises rapidly, becoming predominant after $r \approx 7$~fm and even exceeding $80\%$ at larger $r$.
This demonstrates that the large spatial extension of the density for $^{37}$Mg shown in Fig.~\ref{fig1} primarily stems from the contribution of the valence neutron, which aligns with the classical picture of nuclear halos.
Given also the sizable gap around the Fermi energy illustrated in Fig.~\ref{fig2}, orbital 5 is well decoupled from other more deeply bound orbitals in terms of both energy and spatial distribution.
This naturally leads to a decomposition of the neutron density, wherein the orbitals below and above the Fermi energy respectively contribute to the core and the halo.
The decomposed density for the core closely resembles that of $^{36}$Mg, as seen in Fig.~\ref{fig1}.
As expected, the halo density extends much farther in space than the core and predominantly contributes to the total neutron density far away from the center.
Therefore, a fully microscopic and self-consistent description of the $p$-wave one-neutron halo in $^{37}$Mg is achieved.

\begin{figure}[htbp]
  \centering
  \includegraphics[width=0.4\textwidth]{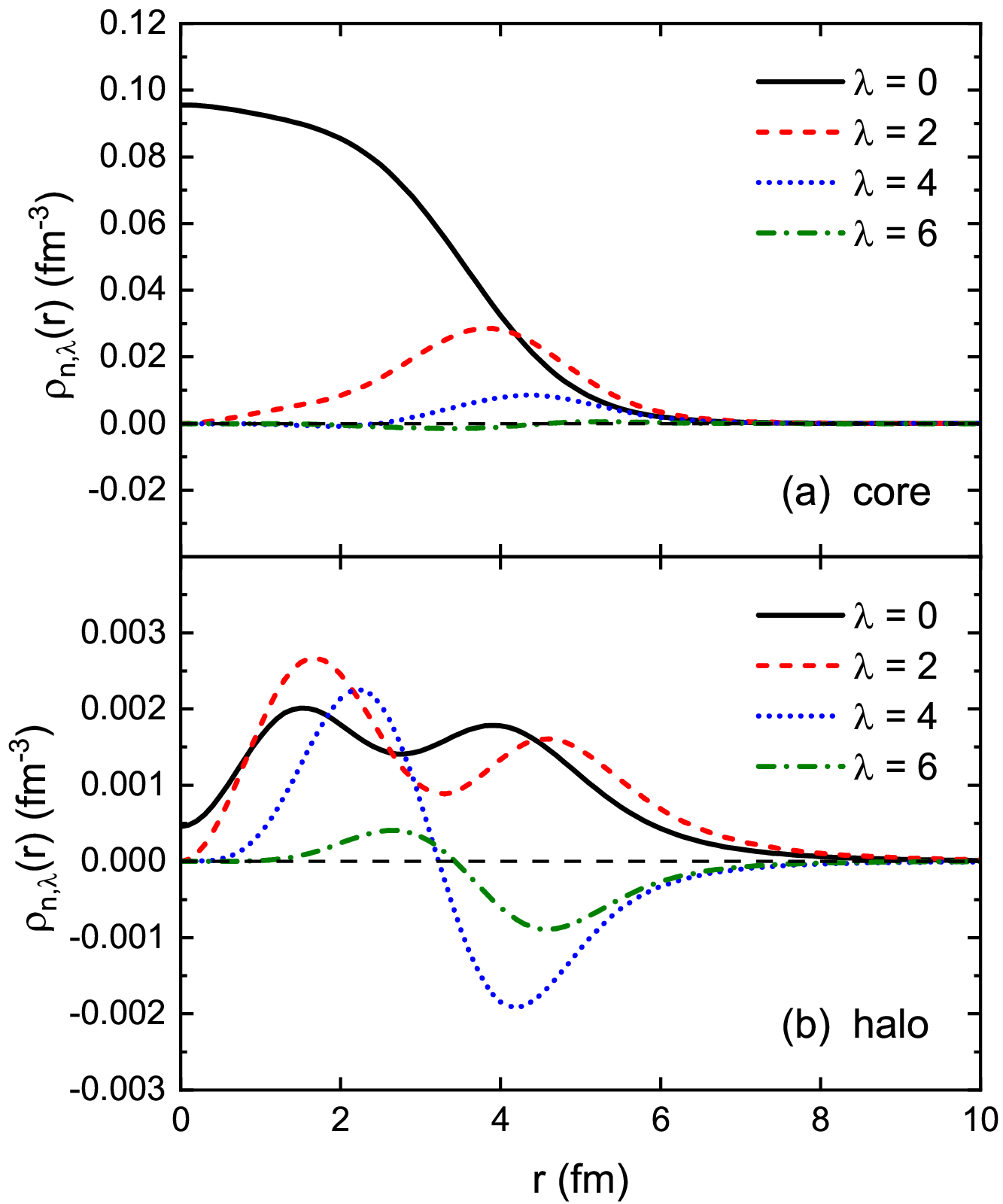}
  \caption{Decomposition of the neutron density into spherical ($\lambda=0$), quadrupole ($\lambda=2$), hexadecapole ($\lambda=4$), and hexacontatetrapole ($\lambda=6$) components for (a) the core and (b) the halo of $^{37}$Mg.}
\label{fig4}
\end{figure}

From Fig.~\ref{fig1}, it can be observed that both the core and the halo of $^{37}$Mg are prolate, indicating that it is a deformed halo nucleus without shape decoupling on the quadrupole level.
However, a recent study based on the triaxial relativistic Hartree-Bogoliubov theory in continuum suggests the possibility of shape decoupling on the triaxial level in deformed halo nuclei \cite{Zhang2022arXiv}.
Therefore, it would be interesting to investigate the higher-order deformations of the core and the halo in $^{37}$Mg.
In Fig.~\ref{fig4}, we decompose the core and halo densities into spherical ($\lambda=0$), quadrupole ($\lambda=2$), hexadecapole ($\lambda=4$), and hexacontatetrapole ($\lambda=6$) components [cf. Eq.~(\ref{legendre})].
The quadrupole components of both the core and the halo are positive, corresponding to their prolate shapes.
In Fig.~\ref{fig4}(a) for the core, the hexadecapole component remains positive, and the hexacontatetrapole component is almost zero, leading to $\beta_4 = 0.23$ and $\beta_6 = 0.01$.
In Fig.~\ref{fig4}(b) for the halo, while the hexadecapole and hexacontatetrapole components are positive when $r\lesssim 3.5$ fm, they become negative for 3.5~fm $\lesssim r \lesssim$ 8~fm.
The radial part of the multipole moment operator, $r^\lambda$, enlarges the contribution of the corresponding component with increasing $r$.
Consequently, the hexadecapole and hexacontatetrapole deformation parameters for the halo are both negative; $\beta_4 = -1.27$ and $\beta_6 = -2.21$.
The presence of differences in hexadecapole and hexacontatetrapole deformations between the core and the halo can be considered as higher-order shape decoupling.
Recent exciting developments in probing nuclear deformation \cite{Giacalone2021PRL,Zhang2022PRL,Bally2022PRL} make it highly anticipated that future experiments can unravel the mystery of shape decoupling in deformed halo nuclei.

Finally, considering the importance of higher-order deformations in superheavy nuclei \cite{Wang2022CPC}, we examine the effects of $\beta_4$ and $\beta_6$ on the ground-state properties of $^{37}$Mg by constraining $\lambda_{\mathrm{max}} = 2$ and $4$ in the calculation.
It turns out that $\beta_4$ and $\beta_6$ influence the binding energy of $^{37}$Mg on the order of 0.1 and 0.01 MeV, respectively, and their impacts on the rms radii are less than 0.01 fm.
For the halo orbital occupied by the valance neutron, considering only the quadrupole deformation $\beta_2$, its single-particle energy $\epsilon$ is $-1.58$ MeV.
The inclusion of $\beta_4$ weakens its binding slightly, with $\epsilon = -1.32$ MeV, and the inclusion of $\beta_6$ marginally changes $\epsilon$ to $-1.34$ MeV.
Particularly, the impacts of $\beta_4$ and $\beta_6$ on the components of the halo orbital are less than $1\%$.
Therefore, the effects of higher-order deformations in light nuclei like $^{37}$Mg are less prominent compared to those in superheavy nuclei.
This does not conflict with the predicted higher-order shape decoupling, as it has been demonstrated that the deformation of a halo depends essentially on the components of the halo orbital(s), irrespective of the shape of the core \cite{Zhou2010PRC(R),Li2012PRC,Sun2018PLB,Zhang2023PRC(L)}.

\section{Summary}

In summary, the deformed $p$-wave halo characteristics of $^{37}$Mg have been described in a microscopic and self-consistent way using the deformed relativistic Hartree-Bogoliubov theory in continuum.
The experimental neutron separation energies and empirical matter radii of neutron-rich magnesium isotopes have been successfully reproduced without any free parameters.
The presence of a neutron halo in $^{37}$Mg is illustrated by the remarkably large spatial extension of its density distribution.
The dominant contribution to the density at large $r$ comes from the valance neutron, which occupies a weakly bound $1/2^-$ orbital with $38\%$ $2p_{1/2}$ and $19\%$ $2p_{3/2}$ components, comparable to the $\approx 40\%$ $p$-wave neutron halo components suggested in experiments.
Furthermore, the consistency of the halo orbital and its components across various employed density functionals, including PC-F1, PC-PK1, NL3*, and PK1, highlights the robustness of the results.
It is regrettable that the DRHBc theory missed predicting $^{37}$Mg as a deformed halo nucleus before 2014.
However, this also presents an opportunity for the DRHBc theory to confidently predict further halo nuclei and other exotic phenomena that may be detected in future experiments.
Despite both the core and the halo of $^{37}$Mg being prolate, higher-order shape decoupling on the hexadecapole and hexacontatetrapole levels has been predicted.
A work using the DRHBc calculated quantities as microscopic inputs in the Glauber model to explore the halo in $^{37}$Mg from the perspective of nuclear reaction is in progress.

\section*{Acknowledgments}

Fruitful discussions with members of the DRHBc Mass Table Collaboration are gratefully appreciated.
This work was partly supported by the National Natural Science Foundation of China (Grant Nos. U2230207, U2030209, 12175010, 11935003, 11975031, 12141501, and 12070131001), the National Key R\&D Program of China (Grant Nos. 2020YFA0406001, 2020YFA0406002, and 2018YFA0404400), and the Institute for Basic Science (IBS-R031-D1).
P.P. was supported by the Rare Isotope Science Project of Institute for Basic Science, funded by the Ministry of Science and ICT MSICT, and National Research Foundation of Korea (2013M7A1A1075764).
M.-H.M. was supported by the National Research Foundation of Korea NRF grants funded by the Korean government Ministry of Science and ICT (Grant Nos. NRF-2021R1F1A1060066, NRF-2020R1A2C3006177, and NRF-2021R1A6A1A03043957).
The results described in this work were obtained using High-performance computing Platform of Peking University.
A portion of the computational resources were provided by the National Supercomputing Center including technical support (No. KSC-2022-CRE-0333).

\section*{References}


\end{document}